\documentclass{emulateapj}
\usepackage{epsf}

\begin{document}

\title{A Robust Test of Evolution near the Tip of the Red Giant Branch and
  Missing Giants in NGC 2808\footnote{Based on observations with the NASA/ESA {\it Hubble Space Telescope},
obtained at the Space Telescope Science Institute, which is operated
by the Association of Universities for Research in Astronomy, Inc.,
under NASA Contract NAS 5-26555.}}

\author{Eric L. Sandquist}

\affil{Department of Astronomy, San Diego State University, 5500 Campanile
Drive, San Diego, CA 92182} 

\email{erics@sciences.sdsu.edu}

\author{Andr\'{e} R. Martel}

\affil{Department of Physics and Astronomy, Johns Hopkins University, 3400
  North Charles Street, Baltimore, MD 21218} 

\email{martel@pha.jhu.edu}

\begin{abstract}
  We describe a new method for robustly testing theoretical predictions of red
  giant evolution near the tip of the giant branch. When theoretical
  cumulative luminosity functions are shifted to align the tip in $I$-band and
  normalized at a luminosity level slightly brighter than the red giant bump,
  virtually all dependence on age and composition (heavy elements and helium
  abundance) is eliminated. While significant comparisons with observations
  require large samples of giant stars, such samples are available for some of
  the most massive Milky Way globular clusters. We present comparisons with
  the clusters NGC 2808 and M5, and find that NGC 2808 has a deficiency of
  bright giants (with a probability of less than about 3\% that a more extreme
  distribution of giant stars would have happened by chance). We discuss the
  possibilities that underestimated neutrino losses or strong mass loss could
  be responsible for the deficit of giants. While we cannot rule out the
  neutrino hypothesis, it cannot explain the apparent agreement between the M5
  observations and models. On the other hand, strong mass loss provides a
  potential link between the giant star observations and NGC 2808's unusually
  blue horizontal branch. If the mass loss hypothesis is true, there is likely
  a significant population of He white dwarfs that could be uncovered
  with slightly deeper UV observations of the cluster.
\end{abstract}


\section{Introduction}

At one level, our understanding of the late stages of the evolution of a
low-mass giant are solid: increasing electron degeneracy in the core coupled
with increasing temperature results in the ``flash'' ignition of helium,
terminating the red giant branch (RGB). However, some of the physics inputs to
the models retain some significant uncertainties \citep{bjork,review}. These
uncertainties (such as electron conduction, neutrino emission, and mass loss)
have significant effects on the observable characteristics of the
brightest stars in old stellar populations, with corresponding influences on
population synthesis models. Testing stellar models for
the late RGB is difficult though because the stars evolve on short timescales 
make up a small fraction of all of the stars in a population.

Because NGC 2808 is a very massive globular cluster ($M_V = -9.39$;
\citealt{harris}) and because the stars within it appear to have nearly
uniform metal content, it provides us with one of the largest ``clean''
samples of stars for examining the late phases in stellar evolution. 
At the same time, NGC 2808 has
an extremely peculiar bimodal distribution of horizontal branch (HB) stars
that has been known since the first moderately-deep photometry was taken
\citep{har74}. The first deep color-magnitude diagram \citep{sosin} was an
even greater shock, revealing a blue HB tail extending 3.5 mag fainter in $V$
and containing two additional gaps in the distribution of stars. As such, NGC
2808 is one of the more obnoxious examples of the ``second parameter''
problem, in which HB star distributions cannot be explained based on
metallicity alone. While some aspects of the HB distribution have been
explained since the original observations, the overall bimodality and extended
blue HB tail have not.
The most recent photometric study of the cluster by
\citet{castell} used archival {\it Hubble Space Telescope (HST)} images
along with wide-field ground-based observations.
Their
sample can be improved upon, however, because they only used a fraction of the
HST observations available. 

\section{Observations and Data Reduction}


Three {\it HST} datasets were used in this work. The only dataset that had not
been previously published was obtained with the HRC detector of the Advanced
Camera for Surveys (ACS) using $F435W$ and $F555W$ filters as part of proposal
ID 10335 (P.I. H. Ford). We also reduced archival WFPC2 images from proposals
6095 and 6804 (PIs Djorgovski and Fusi Pecci). The HRC images mostly overlap
the fields of the PC chips, but allowed us to resolve
a number of stars that were blended in the WFPC2 images.

The HRC frames were processed using the DOLPHOT photometry package
\footnote{http://purcell.as.arizona.edu/$\sim$andy/dolphot/} with its module
tuned for ACS data. 
Individual frames (prior
to drizzling in the ACS pipeline) were obtained from the ACS team in order to
get photometry on the giant stars.
For the
photometry, differences in effective pixel area were corrected within
DOLPHOT through the use of a pixel area map provided on the ACS website
\footnote{http://www.stsci.edu/hst/acs/analysis/PAMS}. 
For
{\it relative} astrometry, we used the fourth-order polynomial corrections
provided in the most recent IDCTAB file for the dataset (q692007bj\_idc.fits)
to facilitate the creation of a common coordinate system with the WFPC2 images.
The WFPC2 images were analyzed using the HSTPhot photometry package
\footnote{http://purcell.as.arizona.edu/$\sim$andy/hstphot/}.
Images from proposal 6095 were taken in the $F218W$, $F439W$, and $F555W$
filters, while those from the proposal 6804 were taken in the $F160BW$,
$F336W$, $F555W$, and $F814W$ filters. $F439W$ and $F336W$ observations were
particularly useful for separating AGB stars from the RGB.
Known geometric distortions in WFPC2 images were
corrected using the METRIC task in the STSDAS package within IRAF. 


Of the filters available, $F814W$ is most appropriate for
giant branch study since near-infrared filters are most directly correlated
with luminosity.
To maximize the size of our sample we used common stars 
from the two WFPC2 fields to derive a transformation from the
$(m_{439} - m_{555},m_{555})$ color-magnitude diagram to $m_{814}$ using a
second-order polynomial containing magnitude and color terms. 
The residuals show no systematic trends with magnitude or
color over the range considered. The median residual was $-0.001$ mag for the
814 stars in common, with a semi-interquartile range (half of the difference
between the 25th and 75th percentile in the ordered list of residuals) of
0.030 mag.



\section{Analysis}

The very
tip in $I$ has been used as a distance indicator for large, old stellar
populations largely thanks to its very weak dependence on age and chemical
composition (see \citealt{dca} for an early reference). This can be seen
empirically in CMDs of composite stellar populations in nearby galaxies (for
recent examples, see \citealt {rizzi,mouh,bella}). Theoretical
predictions confirm that the composition dependence is very weak for [Fe/H]
$\la -1$ \citep{vr}.

\begin{figure}
\plotone{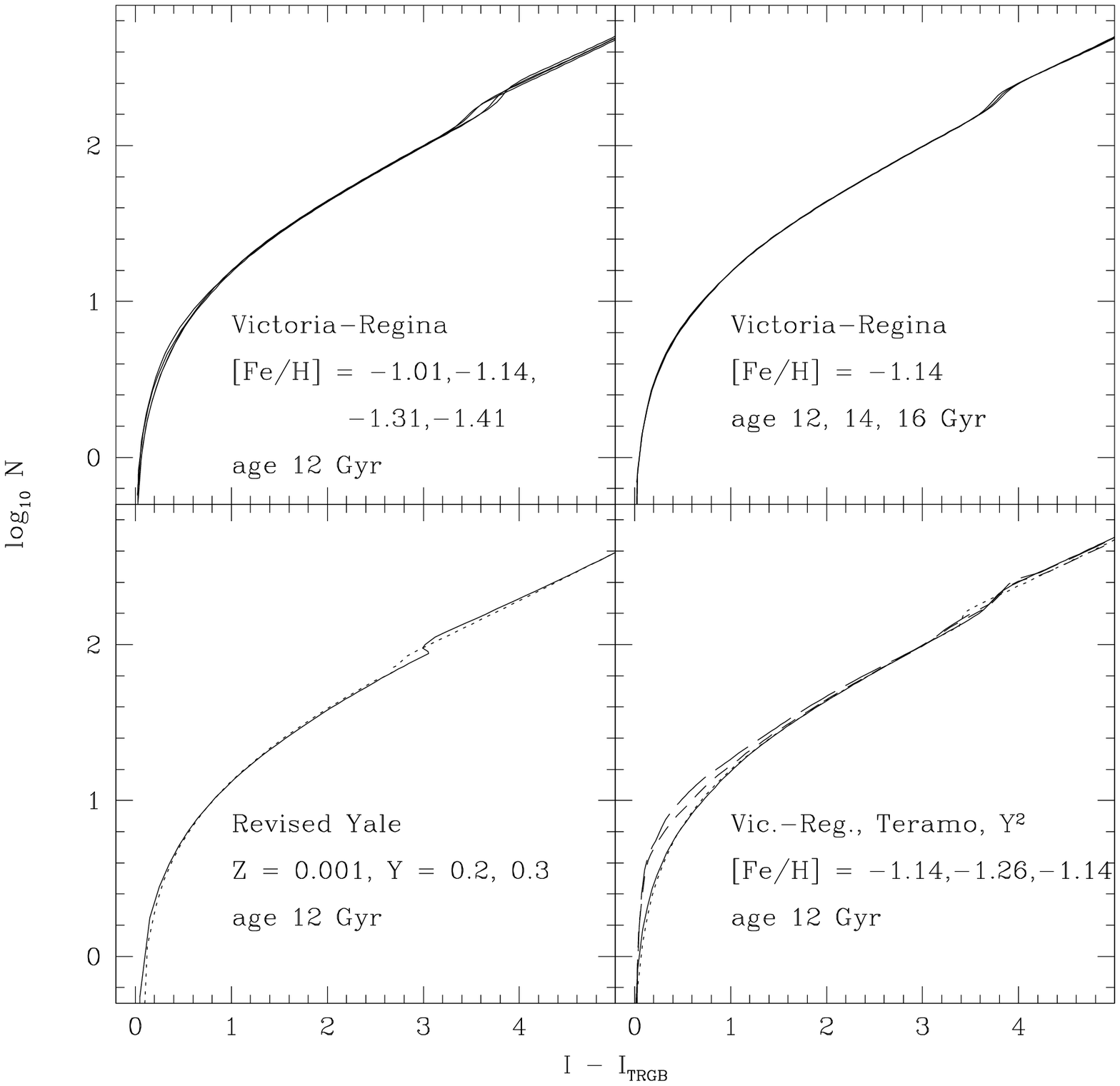}
\caption{Comparisons of theoretical cumulative luminosity functions after
  shifting in magnitude to match the tip of the RGB in $I$ band, and
  normalizing just above the RGB bump. The models are Victoria-Regina
  \citep{vr}, Teramo \citep{ter}, Yale-Yonsei \citep{yy}, and Revised Yale
  \citep{ryi} isochrones. In the lower right panel, the Victoria-Regina and
  Teramo models overlap. The two Yale-Yonsei models with $T_{eff}$-color
  transformations from \citealt{ryi} ({\it short dashed line}) and
  \citealt{lej} ({\it long dashed line}) fall higher.
\label{theory}}
\end{figure}

Even with the large number of RGB stars in NGC 2808, it is
worthwhile to find the most robust comparison with
models possible. The tests below involve cumulative luminosity functions
(LFs), counting stars starting at the observed TRGB.  A comparison with models
requires a magnitude shift (akin to the distance modulus) and a
vertical normalization. For the horizontal (magnitude) shift, we have opted to
shift the models to the observed magnitude of the brightest cluster giant. As
noted above, this has the advantage that it is largely independent of chemical
composition and age (although its absolute position is affected by
uncertainties in physics that affect the timing of the helium flash). For the
vertical normalization, we have forced the models to have a number of stars
equal to the number of observed stars just brighter than the RGB bump at
$m_{814} = 14.8$. 
As seen in Fig. \ref{theory},
by normalizing in this way, the models for a wide range of compositions
(heavy elements and helium) and ages overlie each other to a high degree. 
To state this another way, the late evolution of RGB stars is
virtually independent of input parameters, and results from the strong
correlation between luminosity and mass of the helium core.  

While there is little dependence on age or composition, there are differences
in physics from model set to model set that affect the cumulative LF. In the
bottom right panel of Fig. \ref{theory}, we compare the predictions from
models by the Yale-Yonsei \citep{yy}, Teramo \citep{ter}, and Victoria-Regina
\citep{vr} models. The Teramo and Victoria-Regina predictions overlie each
other almost perfectly, while the Yale-Yonsei models predict considerably more
giants (and therefore slower evolution) near the TRGB.  The two Yale-Yonsei
models use different bolometric corrections.  While this does affect the
cumulative LF, it does not account for the entire difference with the
Victoria-Regina and Teramo models. The implementation of neutrino energy loss
rates is probably responsible for the difference: the Yale-Yonsei models
employ \citet{i89} rates, while the Teramo models use \citep{hfw} rates and
the Victoria-Regina models use \citet{i96} rates.  The dominant plasma
neutrino emission rates have been updated several times since the \citet{i89}
paper, and the values used in the Yale-Yonsei isochrones are probably too low,
which allows giants to evolve more slowly because nuclear reactions don't need
to provide for larger neutrino energy losses. As corroboration, we note that
the Yale-Yonsei models have lower luminosities at the TRGB ($\log (L/L_{\sun})
\approx 3.33$) than the Teramo and Victoria-Regina models ($\log (L/L_{\sun})
\approx 3.37$), consistent with lower cooling rates and an earlier flash.
This is an important issue because systematic uncertainties in the
characteristics of the helium flash translate to corresponding uncertainties
in the properties of HB stars, including their luminosities.

\begin{figure}
\plotone{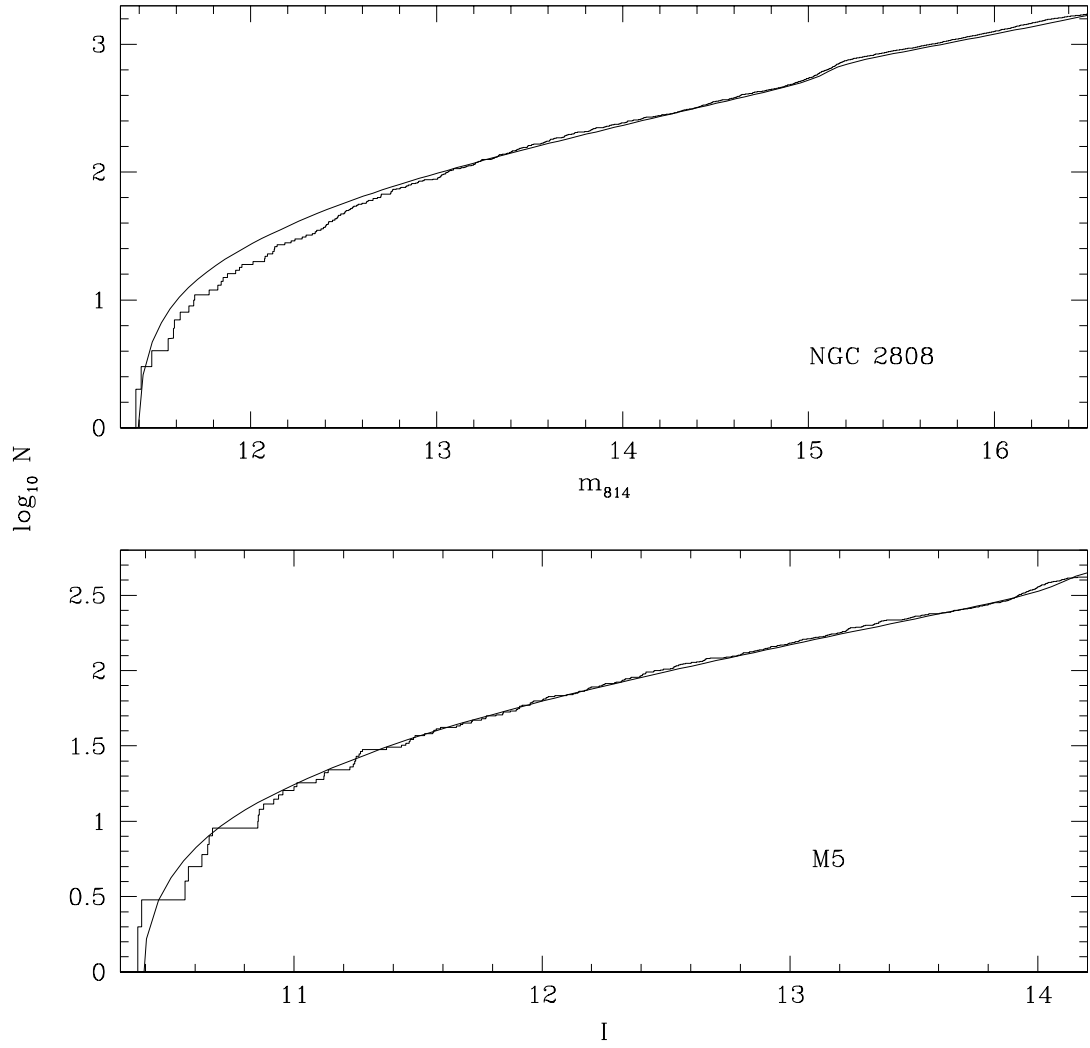}
\caption{The cumulative luminosity function for bright RGB stars for NGC 2808
  (this work) and M5 \citep{sb}, along with a Victoria-Regina model \citep{vr}
  for [Fe/H] = -1.14 and age 12 Gyr.
\label{obs}}
\end{figure}

In Fig. \ref{obs}, we compare the observations for NGC 2808 and for M5
\citep{sb} to the Victoria-Regina models (though the results are nearly the
same for the Teramo models). To judge the significance of the differences, we
used two methods to estimate the probability of detecting a {\it smaller}
sample of bright giants if they were drawn from a cumulative distribution
given by the Victoria-Regina models --- these are essentially estimates of the
probability of a ``false alarm''. First, we did one-sided Kolmogorov-Smirnov
(K-S) tests at various brightness levels.  This was necessary because
adjustment of the faint limit for the giant sample affects the size of an
absolute deviation in the cumulative distributions (the statistical quantity
used in the test) as well as the total sample size, both of which affect the
computed significance of a deviation. When the faint limit is set 1.8
magnitudes below the TRGB, the RGB sample is 142 stars for NGC 2808, the
maximum absolute deviation is 0.11, and the probability that a sample of the
same size with a more extreme deviation would be drawn from the theoretical
distribution is 4.3\%. By comparison, for M5 the sample was 91 stars, the
maximum absolute deviation was 0.06, and the probability was 89\% that a
sample with a more extreme deviation would be drawn from the theoretical
distribution. By comparison, there is a 0.9\% chance that a distribution drawn
from the Yale-Yonsei model would deviate to a larger degree than the observed
NGC 2808 distribution, and a 28\% chance that one would deviate more than the
M5 distribution.

Our second method used predictions from a binomial distribution function.
\citet{sal} used a similar formulation to determine the likelihood of finding
stars within a certain magnitude range near the TRGB. If we have a total
sample of RGB stars $N_{RGB}$ brighter than a certain level, then the
theoretical cumulative luminosity function can be used to predict the
probability $P_i$ that any one star in that sample will be found in the
brighter portion of the sample (brighter than a second faint limit that is
closer to the TRGB). If the theoretical cumulative luminosity function
provides an accurate model of the relative
evolutionary timescales for the giants, then the probability of measuring a
number of stars $n \leq N$ in the brighter portion of the sample is
\[ P_{\leq N} = \sum_{n=1}^{N} \frac{N_{RGB}!}{n! (N_{RGB} - n)!} P_i^n
(1-P_i)^{(N_{RGB} - n)}. \] For NGC 2808, we used the RGB sample between the
TRGB and $m_{814} = 14.8$ just brighter than the RGB bump ($N_{RGB} = 441$).
We find that the probability reaches a minimum of about 1.3\% for a faint
limit of $m_{814} = 12.4$ ($N = 36$ and $P_i = 0.115$), but is less than about
5\% for $m_{814} \leq 12.5$ ($N = 46$). Both of our methods agree that the
probability of a false alarm for NGC 2808 is (conservatively) less than 5\%.

The horizontal (magnitude) shift used to align the model and observed TRGB has
a small effect on the calculated probabilities.  Because we use the brightest
{\it observed} giant, we underestimate the luminosity of the TRGB by some
amount. However, because the models appear to be overestimating the number of
bright giants, this only ends up making our test probabilities conservative
overestimates. The probability of finding at least one star within a certain
magnitude of the TRGB can be determined from the binomial distribution and an
assumed cumulative luminosity function \citep{sal}. Using the Victoria-Regina
models, we find there would be a 50\% probability of having at least one star
within 0.02 mag of the TRGB for a sample of the same size as the one in NGC
2808. If the brightest NGC 2808 giant is as much as 0.05 mag fainter than the
TRGB (the binomial distribution predicts less than a 5\% chance of this), the
K-S probability would be reduced to 2.6\%. The probabilities derived from the
binomial distribution increase slightly if the TRGB brightness is
underestimated: a 0.05 mag underestimate increases the minimum probability
from 1.3\% to 2.1\%.  Thus, the deficit of bright RGB stars in NGC 2808 seems
secure.




We have two possible explanations for the observations. First, larger neutrino
emission rates would accelerate evolution near the TRGB, as mentioned above
with regard to the Yale-Yonsei models. The NGC 2808 observations may be
showing this clearly, while statistical fluctuations may be concealing it in
M5. We have no reason to believe that RGB neutrino emissions should differ
systematically from cluster to cluster, so this explanation would disconnect
the deficit of RGB stars from discussions of HB morphology.  The $I$-band
location of the discrepancy in the NGC 2808 CLF (starting $\sim 1.6$ mag below
the TRGB) holds some physical information if it results from
greater-than-predicted cooling of giant cores.  As long as the temperature and
density sensitivities of the plasma neutrino emission rates agree with those
of \citet{hfw}, modifications of the emission rate are unlikely to explain the
NGC 2808 observations because they would primarily affect RGB stars within
about 1 mag of the TRGB.  For typical RGB core conditions, the energy loss
rate has fairly large temperature and density dependences ($q \propto
\rho^{2.5}T^{9}$, evaluated from the \citealt{hfw} formulae) that are
responsible for a rapid increase in energy loss.  Of course, if there is an
energy loss mechanism having different density and temperature dependences,
then the shape of the LF would be changed. While we cannot rule out the
possibility of systematic errors in the theoretical neutrino loss rates used
in the stellar models, new physics would probably be required to match the
observations.

A second explanation involves enhanced mass loss near the TRGB for a {\it
  fraction} of the stars. Models
predict that stars can leave the RGB before He flash if the mass of the star's
envelope decreases below a critical value. These stars can have a ``hot
flash'' He ignition after leaving the RGB, before finally appearing on the
blue end of the HB for a time approximately as long as that of a typical HB
star. With even more mass loss on the RGB, a star will leave the RGB earlier,
a hot He flash will be prevented, and a He white dwarf would result.
According to models \citep{dcruz}, stars leaving the RGB before they get
within 0.4 mag of the TRGB will produce He white dwarfs.
In NGC 2808, the vast majority of the stars seen on the HB in NGC 2808 should
have gone through the entire RGB phase (complete with He flash) because their
positions on the HB imply substantial envelope mass.  However, the deficit of
stars more than 0.4 mag fainter than the TRGB would imply we should expect to
have a significant population of He core white dwarfs if the mass loss
hypothesis is true. Based on the lack of a blue HB tail in M5, we would not
expect an RGB deficit there.

The population ratio $R = N_{HB} / N_{RGB} = 1.62\pm0.07$ \citep{castell} is
near the average for Galactic globular clusters \citep{saly}. However, this
ratio could retain a ``normal'' value if giants are removed exclusively from
the upper RGB (which is the most sparsely populated), or if the stars avoiding
the HB phase (by never igniting He) are small in number.  If there is a large
population of He white dwarfs, they may be detectable at the bright end of the
white dwarf cooling sequence because they cool more slowly than the CO white
dwarfs produced by traditional evolutionary paths \citep{castell2}.
\citet{dieball} identified about 40 white dwarf candidates using STIS UV
imagery, and found the numbers to be in rough agreement with theoretical
predictions. Unfortunately, the \citeauthor{dieball} survey does not reach far
down the WD sequence, and models \citep{seren} predict that the cooling age of
a He WD is only about twice the cooling age of a CO WD near their completeness
limit. (The ratio becomes much more extreme the fainter one goes.) Thus, if
the He WD population is less than the size of the observed population, the
production rate for He WD must be less than about half the CO WD production
rate.

If the true luminosity function of RGB stars is given by the Victoria-Regina
or Teramo models, then NGC 2808 shows a deficit in $\log N$ of at most about
0.1 (20\% in $N$) in the magnitude range where He WDs are probably produced
($I - I_{TRGB} \ga 0.4$). If these ``missing'' giants produce He WDs directly,
then this would enhance the observable population of WDs in the
\citeauthor{dieball} sample by about 40\%. The uncertainties in the
theoretical predictions and in the WD numbers are not yet able to
rule this possibility out. A slightly deeper survey of the cluster's WD
population would help definitively settle whether there is a significant
population of He WDs.  Such a survey has been done for $\omega$ Cen
\citep{monelli}, another cluster with a extensive blue HB tail, and again the
results are in rough agreement with evolutionary timescales. However,
\citeauthor{monelli} also note that the observed WDs appear to be redder than
expected, and one of the possible explanations of this is a large
population of He core WDs.





\section{Discussion}


Because our analysis implies that the RGB evolution is virtually independent
of chemical composition and age inputs, it should be possible to merge
photometric data for stars from clusters with relatively heterogeneous
characteristics. Multiple cluster samples could place {\it very} tight
constraints on physics inputs like neutrino losses (and non-standard neutrino
emission mechanisms as well).
Indeed, our comparisons indicate that the neutrino losses used by the
Yale-Yonsei models can already be ruled out based on the comparisons with the
massive globular clusters M5 and NGC 2808.

Although our method can only be applied in a
practical way to the most massive globular clusters, it does provide a new way
of probing the ``second parameter'' problem in horizontal branch stars. If
there is a dynamical influence on cluster giants causing them to leave the RGB
early in NGC 2808, then we would expect similar features to be present in
other clusters with extreme blue HB tails. Alternately, for a cluster like 47
Tuc with no obvious blue HB extension, the upper RGB should match models. 


\acknowledgments We would like to thank H. Ford for sharing the ACS images
from his HST program 10335, A. Dolphin for clarifications of aspects of the
DOLPHOT software, and C.-D. Lin for statistical advice on this paper. This
work has been funded through grant 05-07785 from the National Science
Foundation to E.L.S. and M. Bolte.


\clearpage

\clearpage

\end{document}